\documentclass[final,5p,times,twocolumn]{elsarticle}
\usepackage{graphicx}
\usepackage{amssymb}
\usepackage{amsmath}
\usepackage{mathtools, cuted}
\usepackage{color}
\usepackage{hyperref}
\usepackage{verbatim}
\usepackage{verbatimbox}
\usepackage{appendix}
\usepackage{float}
\usepackage{listings}
\usepackage{xcolor}
\usepackage[most]{tcolorbox}
\usepackage{pifont}% symbols
\usepackage{upquote}

\usepackage{array}
\usepackage{booktabs}
\usepackage{threeparttable}
\usepackage{multirow}
\usepackage{makecell}
\usepackage{arydshln}
\hypersetup{colorlinks = true,linkcolor = blue,anchorcolor =red,citecolor = blue,filecolor = red,urlcolor = red, pdfauthor=author}

\definecolor{codegreen}{rgb}{0,0.6,0}
\definecolor{codegray}{rgb}{0.5,0.5,0.5}
\definecolor{codepurple}{rgb}{0.58,0,0.82}

\lstdefinelanguage{QE}{
    keywords={CONTROL,SYSTEM,ELECTRONS,ATOMIC_SPECIES,ATOMIC_POSITIONS,K_POINTS,IONS,CELL,INPUTPP,INPUTPH,PLOT,ENERGY_GRID,INPUT},
    morecomment=[l]{\#}
}

\lstdefinelanguage{Command}{
    keywords={cd,ls,mkdir,mv,rm,cp,pw.x,mpirun,ph.x,tail,vi,for,xcrysden,grep},
    morecomment=[l]{\#}
}

\newtcblisting{codepython}[2][]{breakable, enhanced, %beamer,
  arc=3pt, top=0mm, bottom=0mm, coltitle=black, colframe=gray!30!white,
  fonttitle=\bfseries, listing only, drop fuzzy shadow=gray!25!white,
  listing options={basicstyle=\footnotesize\ttfamily,language=Python,upquote=true,showstringspaces=false,commentstyle=\color{blue}},
  title=#2}

\newtcblisting{codeQE}[2][]{breakable, enhanced, %beamer,
  arc=3pt, top=0mm, bottom=0mm, coltitle=black, colframe=gray!30!white,
  fonttitle=\bfseries, listing only, drop fuzzy shadow=gray!25!white,
  listing options={basicstyle=\footnotesize\ttfamily,language=QE,upquote=true,showstringspaces=false,commentstyle=\color{blue}},
  title=#2}

\newtcblisting{codeQEsmall}[2][]{breakable, enhanced, %beamer,
  arc=3pt, top=0mm, bottom=0mm, coltitle=black, colframe=gray!30!white,
  fonttitle=\bfseries, listing only, drop fuzzy shadow=gray!25!white,
  listing options={basicstyle=\scriptsize\ttfamily,language=QE,upquote=true,showstringspaces=false,commentstyle=\color{black}},title=#2}

\newtcblisting{terminal}[2][]{breakable, enhanced, arc=3pt, top=0mm,
  bottom=0mm, left=-7mm, right=-2mm, coltitle=black, colframe=gray!30!white,
  fonttitle=\bfseries, listing only, drop fuzzy shadow=gray!25!white,
  listing options={numbers=none,frame=none,basicstyle=\ttfamily,language=Bash,upquote=true,showstringspaces=false,escapeinside={*}{*}},
  title=#2}

\newtcblisting{terminal2}[2][]{breakable, enhanced, arc=3pt, top=0mm,
  bottom=0mm, left=2mm, right=0mm, coltitle=black, colframe=gray!30!white,
  fonttitle=\bfseries, listing only, drop fuzzy shadow=gray!25!white,
  listing options={numbers=none,frame=none,basicstyle=\scriptsize\ttfamily,language=Bash,upquote=true,showstringspaces=false,escapeinside={*}{*}},
  title=#2}
  

\definecolor{codegreen}{rgb}{0,0.6,0}
\definecolor{codegray}{rgb}{0.5,0.5,0.5}
\definecolor{codepurple}{rgb}{0.58,0,0.82}
\definecolor{backcolour}{rgb}{0.95,0.95,0.92}
\journal{Computer Physics Communications}

\begin{document}

\begin{frontmatter}

\title{{\sc QERaman}: An open-source program for calculating resonance Raman spectra\\ based on {\sc Quantum ESPRESSO}}

\author[a,b]{Nguyen T. Hung\corref{author}}
\author[c]{Jianqi Huang}
\author[b]{Yuki Tatsumi}
\author[c]{Teng Yang}
\author[b]{Riichiro Saito\corref{author}}
%%Please add your name to the author list according to your preferred choice. 

\cortext[author] {Corresponding authors.\\\textit{E-mail address:} nguyen.tuan.hung.e4@tohoku.ac.jp; r.saito.sendai@gmail.com}
\address[a]{Frontier Research Institute for Interdisciplinary Sciences, Tohoku University, Sendai 980-8578, Japan}
\address[b]{Department of Physics, Tohoku University, Sendai 980-8578, Japan}
\address[c]{Shenyang National Laboratory for Materials Science, Institute of Metal Research, Chinese Academy of Sciences, School of Material Science and Engineering, University of Science and Technology of China, Shenyang 110016, P. R. China}

%ABSTRACT
\begin{abstract}
We present an open-source program {\sc QERaman} that computes first-order resonance Raman spectroscopy of materials using the output data from {\sc Quantum ESPRESSO}. Complex values of Raman tensors are calculated based on the quantum description of the Raman scattering from calculations of electron-photon and electron-phonon matrix elements, which are obtained by using the modified {\sc Quantum ESPRESSO}. Our program also calculates the resonant Raman spectra as a function of incident laser energy for linearly- or circularly-polarized light. Hands-on tutorials for graphene and MoS$_2$ are given to show how to run {\sc QERaman}. All codes, examples, and scripts are available on the GitHub repository.
\end{abstract}


\end{frontmatter}
{\bf PROGRAM SUMMARY}

\begin{small}
\noindent
{\em Program Title:} {\sc QERaman} \\
{\em Developer's repository link:} \href{https://github.com/nguyen-group/QERaman}{https://github.com/nguyen-group/QERaman}\\
{\em Licensing provisions:} GNU General Public Licence 3.0\\
{\em Programming language:} Fortran\\
{\em External routines}: {\sc Quantum ESPRESSO v7.2}\\
{\em Nature of problem:}  Resonance Raman spectra with first-principles calculations. \\
{\em Solution method:} The Raman intensity formula is given by the quantum theory, in which the electron-photon and the electron-phonon matrix elements are obtained from the modified {\sc Quantum ESPRESSO} package.
\end{small}

%% main text
\section{Introduction}\label{sec:intro}
Raman spectroscopy is the inelastic scattering of light in a material, which is one of the standard tools to characterize the   material~\cite{dresselhaus2005raman,jorio2011raman,saito2016raman}. In Raman spectra, we plot the intensity of the scattered photon as a function of the energy shift from the incident light (Raman shift in units of cm$^{-1}$). Raman spectroscopy is a non-destructive method measured at room temperature and ambient pressure without complicated sample preparation~\cite{jorio2011raman}. Micro Raman spectroscopy has recently been frequently used for low-dimensional materials, such as one-dimensional or two-dimensional (1D or 2D) materials, for characterizing the structure, electronic, or phonon properties in the optical microscope~\cite{jimenez2000micro,jorio2001structural,zhang2019thermal}. 

In the low-dimensional materials, the joint density of state (JDOS) has a singularity at specific energies, known as van Hove singularity (VHS), at which we observed a strong, resonant Raman signal~\cite{dresselhaus2005raman,dresselhaus2002raman,saito2011raman}. In the resonant Raman spectra, the Raman intensity is enhanced significantly (say, 1000 times) compared with the non-resonant Raman signal when we match either incident or scattered photon energy to the VHS energy, which we call the resonance Raman effect~\cite{jorio2001structural}. When we have several laser energies for the incident light, we can compare the Raman spectra from which we can know not only the resonant effect but also symmetry of the unoccupied electronic states by comparing them with the calculated Raman spectra. When we measure Raman intensity for a particular phonon mode as a function of laser energy, which we call the Raman excitation profile, we can obtain information on the VHS in electronic energies~\cite{zhang2020anomalous,zhang2020anisotropic}. In the Raman excitation profile, the width of the peak corresponds to the inverse of the lifetime of photo-excited electrons, while the spectra width of Raman spectra for a given phonon mode corresponds to the inverse of the lifetime of the phonon. Thus resonance Raman spectra give a lot of information compared with non-resonant Raman spectra. However, in the conventional first-principles package, we can calculate only non-resonant Raman calculation. The goal of the paper is to give the open-source of resonant Raman spectra within the first-principles package.

Using the calculated Raman tensor, whose elements are complex values, we can analyze the symmetry of phonon modes by helicity-dependent Raman spectra using circularly polarized light (CPL)~\cite{tatsumi2018conservation,tatsumi2018interplay}. The calculated results can be directly compared with the resonant Raman spectra by using a possible combination of helicity (left or right) of the CPL in the incident and scattered light. When we use CPL for incident light, the helicity of the scattered CPL either changes or conserves from that of the incident CPL depending on the symmetry of the phonon mode. For example, the out-of-plane $A$ mode of the 2D MoS$_2$ is the helicity-conversed Raman mode, while the in-plane $E$ mode is the helicity-changing Raman mode~\cite{tatsumi2018interplay}. However, though the origin of the helicity-dependent Raman mode can be explained by the Raman tensor, each element of the Raman tensor can not be directly observed by the experiment~\cite{han2022complex}. Therefore, calculating the Raman tensor and resonance Raman spectra is essential to understand the measured Raman spectra. 

In order to obtain the complex Raman tensor, the complex values of electron-photon matrix elements and electron-phonon matrix elements are required from the first-principles calculations. {\sc Quantum ESPRESSO (QE)} is a free, open-source package for the first-principles calculation using plane-wave basis sets and pseudopotentials~\cite{giannozzi09-espresso}. In the {\sc QE} package, the Raman tensor can be calculated by \verb|dynmat.x| code~\cite{lazzeri2003first}. However, the Raman tensor calculation from \verb|dynmat.x| is the non-resonance Raman process using the Placzek approximation and supports only the non-metal system within the local-density approximations (LDA)~\cite{hung2022quantum}. We thus develop an open-source code, named {\sc QERaman}, that can calculate the resonance Raman process for metallic or semiconducting materials, in which the electron-photon matrix elements and electron-phonon matrix elements are obtained from the modified \verb|bands.x| and \verb|ph.x| codes in the {\sc QE} package, respectively. The {\sc QERaman} code allows everyone to compute the resonance Raman spectra by specifying the laser energy, which can be analyzed even by experimentalists. We also plan to expand the present result to double-resonance Raman spectra in the near future~\cite{huang2022first}. 

The structure of this paper is organized as follows. In Sec.~\ref{sec:raman}, we briefly introduce the theoretical background of the resonance Raman spectroscopy from the quantum description. In Sec.~\ref{sec:install}, we explain the installation and usage of {\sc QERaman}. After providing some examples: graphene and MoS$_2$ in Sec.~\ref{sec:example}, a summary is given in Sec.~\ref{sec:summary}.

\section{Quantum description of the Raman scattering}
\label{sec:raman}

\begin{figure}[t]
  \centering \includegraphics[clip,width=6cm]{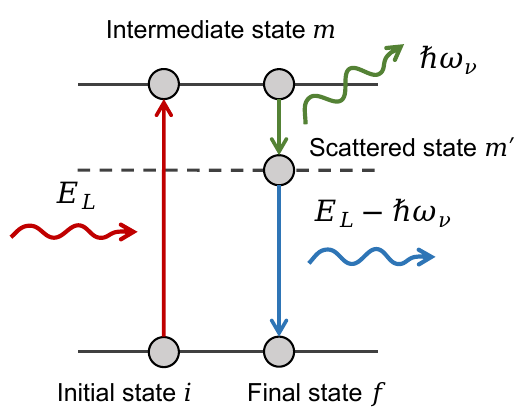}
  \caption{\label{fig:raman} Raman scattering process, in which $E_L$ is laser energy of optical light source, and $\hbar\omega_\nu$ is the phonon frequency at $\nu$ mode.}
\end{figure}

In the time-dependent perturbation theory of quantum mechanics, the first-order Raman scattering follows the three sub-process, as shown in Fig.~\ref{fig:raman}, that is: (1) the electron in the initial state $i$ gets the laser energy $E_L$ of the incident photon and excites to the intermediate state $m$, (2) the electron at the $m$ state interacts with the phonon and emit (or absorb) phonon energy of $\hbar \omega_\nu$, where $\hbar$ is the reduced Planck's constant and $\omega_\nu$ is the phonon frequency of the $\nu$ mode at the $\Gamma$ point ($\mathbf{q}=0$), to go to the scattered state $m^{\prime}$, and (3) the electron at the $m^{\prime}$ state recombines with the hole and emits the scattered photon with an energy of $E_L - \hbar \omega_\nu$ to go to the final state $f$. It is noted that the $f$ state should be identical to the $i$ state (i.e., $f=i$) for the electron to recombine with a hole. In the first-order Raman scattering, only the phonon at the $\Gamma$ point is emitted or absorbed. The first-order Raman intensity formula as a function of incident laser energy $E_L$ and Raman shift $E_{RS}$ is given by the third-order perturbation theory~\cite{jorio2011raman} as follows:
\begin{strip}
\begin{equation}
    \label{eq:1}
    I(E_L, E_{RS}) = \sum_{\nu}\left|\sum_{\mathbf{k}}\sum_{i,m,m^{\prime}}\frac{
    \mathcal{M}_{\text{opt}}^{m^{\prime} \to i}(\mathbf{k})
    \mathcal{M}_{\text{ep}}^{m \to m^{\prime}}(\mathbf{k},\nu)
    \mathcal{M}_{\text{opt}}^{i \to m}(\mathbf{k})
    }{[E_L-\Delta E_{mi}(\mathbf{k})][E_L-\Delta E_{m^{\prime}i}(\mathbf{k})-\hbar\omega_\nu]}\right|^2 \delta(E_{RS}-\hbar\omega_\nu),
\end{equation}
\end{strip}
where $\Delta E_{m(m^{\prime})i}(\mathbf{k})=E_{m(m^\prime)}(\mathbf{k})-E_{i}(\mathbf{k})-i\gamma$ is the energy difference between $m (m^{\prime})$ and $i$ states at the wavevector $\mathbf{k}$ of an electron with a resonance window $\gamma$, which is related to the lifetime of the photoexcited carrier. In the Raman spectra, the Raman intensity is plotted as a function of $E_{RS}$, in which intensity has a value only at each $E_{RS}=\hbar\omega_{\nu}$. To compare the calculated Raman spectra with the experiment with a finite width for each spectra peak, we approximate the delta function to the Lorentzian function with the spectral width $\Gamma$ as 
\begin{equation}
    \label{eq:2}
    \delta(E_{RS}-\hbar\omega_\nu) \to \frac{1}{\pi}\left[\frac{\Gamma}{(E_{RS}-\hbar\omega_\nu)^2+\Gamma^2}\right],
\end{equation}
where $\Gamma$ is proportional to the inverse of the lifetime of the phonon. $\mathcal{M}_{\text{opt}}^{m^{\prime} \to i}(\mathbf{k})$ is the electron-photon matrix elements between $m^{\prime}$ and $i$ states, as explained in \ref{sec:elpt}, and $\mathcal{M}_{\text{ep}}^{m \to m^{\prime}}(\mathbf{k},\nu)$ is the electron-phonon matrix elements between $m$ and $m^{\prime}$ states, as explained in \ref{sec:elph}. From Eq.~\eqref{eq:elpt14}, $\mathcal{M}_{\text{opt}}^{m^{\prime} \to i}(\mathbf{k})$ is expressed within the dipole approximation~\cite{gruneis2003inhomogeneous}  by:
\begin{equation}
    \label{eq:3}
\mathcal{M}_{\text{opt}}^{m^{\prime} \to i}(\mathbf{k})= C_{\text{opt}} \mathcal{D}^{m^{\prime} \to i}(\mathbf{k})\cdot \mathbf{P},
\end{equation}
where $C_{\text{opt}}$ is the potential associated with light (Eq.~\eqref{eq:elpt14a}), $\mathcal{D}^{m^{\prime} \to i}(\mathbf{k})=\langle \psi^{i}_\mathbf{k}|\nabla| \psi^{m^{\prime}}_\mathbf{k} \rangle$ is the dipole vector (Eq.~\eqref{eq:elpt15}), and $\mathbf{P}$ is the polarization vector (Eq.~\eqref{eq:elpt13}). By inserting Eq.~\eqref{eq:3} into Eq.~\eqref{eq:1}, the Raman intensity formula can be written as
\begin{equation}
    \label{eq:4}
    I_{\sigma^{\prime},\sigma}(E_L, E_{RS}) = \sum_{\nu}\left|\mathbf{P}_{\sigma^{\prime}}^*\mathbf{R}(\nu)\mathbf{P}_{\sigma}\right|^2 
    \delta(E_{RS}-\hbar\omega_\nu),
\end{equation}
where $\mathbf{R}(\nu)$ is the Raman tensor, which is given by
\begin{equation}
    \label{eq:5}
    \mathbf{R}(\nu) = C_{\text{opt}}^2 \sum_{\mathbf{k}}\sum_{i,m,m^{\prime}}\frac{
    \mathcal{D}^{m^{\prime} \to i}(\mathbf{k})
    \mathcal{M}_{\text{ep}}^{m \to m^{\prime}}(\mathbf{k},\nu)
    \mathcal{D}^{i \to m}(\mathbf{k})
    }{[E_L-\Delta E_{mi}(\mathbf{k})][E_L-\Delta E_{m^{\prime}i}(\mathbf{k})-\hbar\omega_\nu]}.
\end{equation}
Here, we assume that the electronic wave number $\mathbf{k}$ of the initial and intermediate states
are much larger than the wave number of the light $\mathbf{k}_{\text{opt}}$ (i.e., $|\mathbf{k}|\gg|\mathbf{k}_{\text{opt}}|$). In Eq. \ref{eq:5}, $C_{\text{opt}}^2$ can be taken out of the summation as a constant in Eq.~\eqref{eq:5}, which is known as the dipole approximation~\cite{jiang2007exciton,gruneis2003inhomogeneous}. For the sake of simplicity, $C_{\text{opt}}^2$ is adopted as a unit in {\sc QERaman} code. $\mathbf{P}_{\sigma^{\prime}}$ and $\mathbf{P}_{\sigma}$ are the polarization vectors (or the Jones vector) of scattered and incident lights, respectively, as explained by Eq.~\eqref{eq:elpt13}. For circularly polarized light with $P_x=P_y$, $\phi={\pi}/{2}$ and $\phi=-{\pi}/{2}$ correspond to left- ($\sigma+$) and right-handed ($\sigma-$) circularly polarized light. From Eq.~\eqref{eq:elpt13}, the polarization vectors for $\sigma+$ and $\sigma-$ is given by
\begin{equation}
    \label{eq:6}
    \mathbf{P}_{\sigma +}=\frac{1}{\sqrt{2}}\left(
    \begin{array}{c}
      1 \\
      i \\
      0
    \end{array}
    \right),~~\text{and} ~~\mathbf{P}_{\sigma -}=\frac{1}{\sqrt{2}}\left(
    \begin{array}{c}
      1 \\
      -i \\
      0
    \end{array}
    \right),
\end{equation}
respectively. By inserting Eq.~\eqref{eq:6} into Eq.~\eqref{eq:4}, we can obtain the helicity-dependent Raman spectra, as shown in the hands-on tutorial for MoS$_2$ monolayer in Sec.~\ref{sec:example}. The Raman intensity in Eq.~\eqref{eq:4} and the Raman tensor in Eq.~\eqref{eq:5} are obtained by using the \verb|raman.x| in the {\sc QERaman} code. 

\section{Installation, workflow, and usage}
\label{sec:installandrun}
In this section, we explain how to install and use {\sc QERaman}. The workflow of the calculations is shown in Fig.2.

\subsection{Download and installation}
\label{sec:install}

\begin{table*}[ht!]
\centering
 \begin{tabular}{p{2.5cm} p{14cm}} 
 \toprule
 \textbf{Command} & \textbf{Purpose}\\ \midrule
 \texttt{bands\_mat.x} & modified \texttt{bands.x} of {\sc QE} to obtain the electron-photon matrix elements (see \ref{sec:elpt}).\\ \hdashline
 \texttt{ph\_mat.x} & modified \texttt{ph.x} of {\sc QE} to obtain the electron-phonon matrix elements (see \ref{sec:elph}).\\ \hdashline
 \texttt{raman.x} & calculate the resonance Raman intensity (see Eq.~\eqref{eq:1}).\\
 \bottomrule
 \end{tabular}
\caption{The commands in {\sc QERaman}.}
\label{tab:command}
\end{table*}

The {\sc QERaman} code is designed as a module of the {\sc QE} distribution, and it resides in a self-contained directory \verb|QERaman| under the root directory of the {\sc QE} tree. The installing environment of {\sc QERaman} (compilers, libraries, etc.) is the same for all modules in the {\sc QE} package. Thus, {\sc QE} must be installed in advance. Then, {\sc QERaman} is installed by following steps:
\begin{enumerate}
    \item[(1)] Download the {\sc QERaman} code for the last stable release at GitHub page:
    \href{https://github.com/nguyen-group/QERaman/releases}{https://github.com/nguyen-group/QERaman/releases}.
    Alternatively, the readers can download the last version of {\sc QERaman} directly from the GitHub repository by following the git command: 
    \begin{terminal}{}
    *\$* git clone https://github.com/nguyen-group/QERaman.git
    \end{terminal}
    The code contains source files and examples in \verb|src| and \verb|examples| folders in the parent directory of 
    \verb|QERaman|, respectively.
    \item[(2)] After downloading the latest version and putting the directory \verb|QERaman| inside the main {\sc QE} directory, the readers can install {\sc QERaman} by following the command: 
    \begin{terminal}{}
    *\$* cd QERaman/src
    *\$* make all
    \end{terminal}
    If everything is done smoothly, three executable files named \verb|bands_mat.x|, \verb|ph_mat.x|, and \verb|raman.x| will be created in the \verb|bin| folder. We list these executable files in Table~\ref{tab:command}. We note that \verb|Makefile| in \verb|QERaman/src| is linked to the libraries of PW, PP, and PH modules of {\sc QE}. Therefore, the reader must install these modules of {\sc QE} before installing {\sc QERaman}. 
\end{enumerate}

\begin{figure}[t]
  \centering \includegraphics[width=8cm]{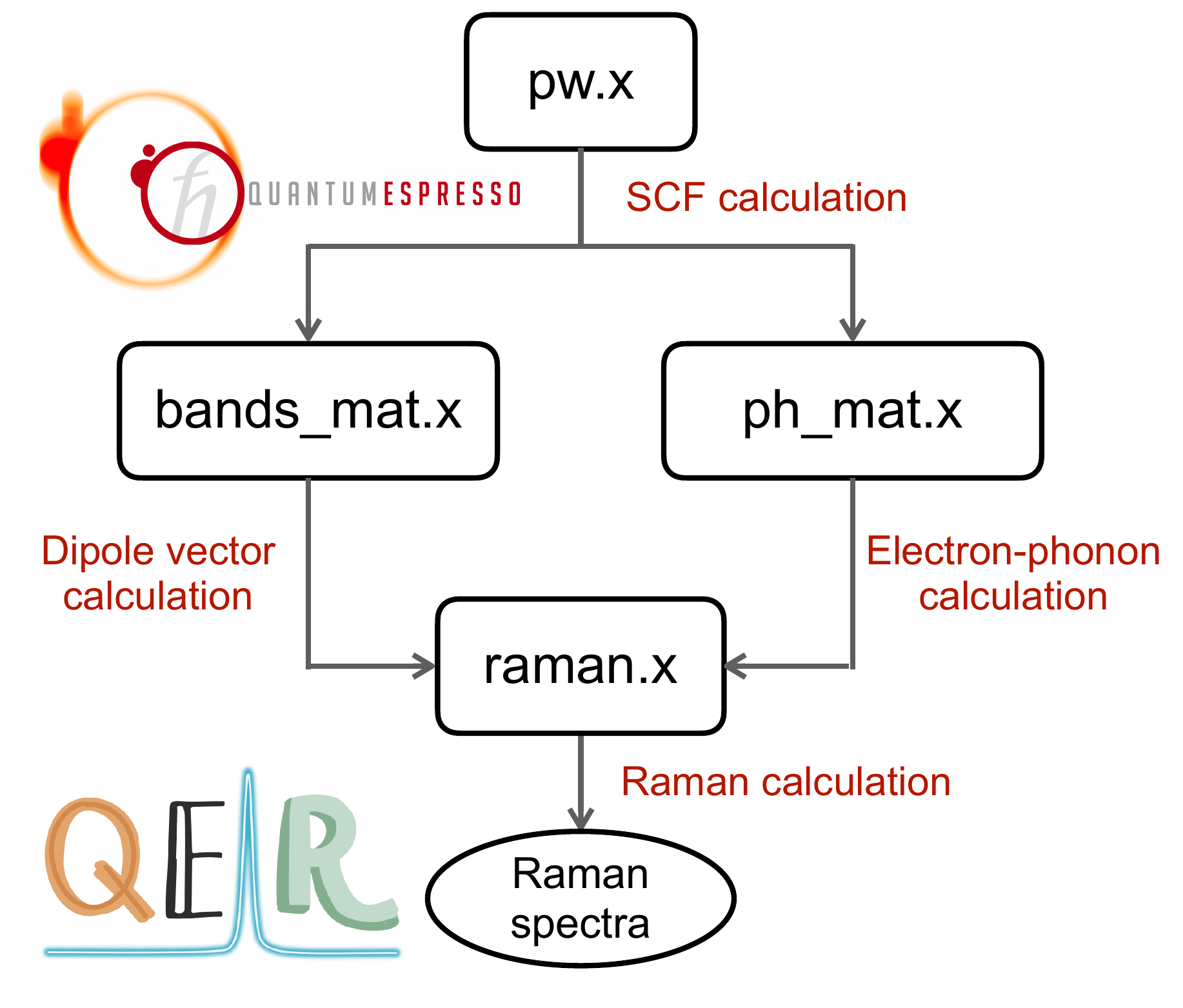}
  \caption{\label{fig:qeraman} Flowchart of the process to calculate the first-order resonance Raman spectra. Four steps are required for the calculation of the resonance Raman spectra. In the first step, the band structure and the wavefunctions of the material are computed by the self-consistent field (SCF) calculation with \texttt{pw.x} of {\sc QE}. In the second step, the dipole vector (see Eq.~\eqref{eq:elpt15}) is calculated by \texttt{bands\_mat.x} of {\sc QERaman}. In the third step, the electron-phonon matrix elements (see Eq.~\eqref{eq:elph1}) is calculated by \texttt{ph\_mat.x} of {\sc QERaman}. Finally, the complex Raman tensor (see Eq.~\eqref{eq:5}) and the Raman spectra (see Eq.~\eqref{eq:4}) are calculated by \texttt{raman.x} of {\sc QERaman}. The top and bottom figures are the logo of {\sc QE} and {\sc QERaman}, respectively.}
\end{figure}

\subsection{Workflow and usage}
\label{sec:run}
In Fig.~\ref{fig:qeraman}, we show the flowchart of the process to calculate the first-order resonance Raman spectra $I(E_L, E_{RS})$. Four steps are required for the calculation of $I(E_L, E_{RS})$ with the commands as follows: 
\begin{terminal2}{}
*\$* mpirun -np 48 pw.x <scf.in> scf.out
*\$* mpirun -np 48 bands_mat.x <bands.in> bands.out
*\$* mpirun -np 48 ph_mat.x <ph.in> ph.out
*\$* raman.x <raman.in> raman.out
\end{terminal2}
\begin{itemize}
    \item[-] \textbf{Step 1}: Run \verb|pw.x| of {\sc QE} with 48 processors (\verb|-np 48|) by the command \verb|mpirun| (running a program with parallel processors) with the input file \verb|scf.in| and the output file \verb|scf.out|. In step 1, the band structure $E(\mathbf{k})$ and the wavefunctions $\psi_\mathbf{k}$ are computed by the self-consistent field (SCF) calculation with the \verb|pw.x| command of {\sc QE}. 
    \item[-] \textbf{Step 2}: Run \verb|bands_mat.x| of {\sc QERaman} with 48 processors with the input file \verb|bands.in| and the output file \verb|bands.out|. In step 2, the dipole vector $\mathcal{D}^{i \to f}(\mathbf{k})$ (see Eq.~\eqref{eq:elpt15}) is calculated using the calculated $\psi_\mathbf{k}$ in the step 1. 
    \item[-] \textbf{Step 3}: Run \verb|ph_mat.x| of {\sc QERaman} with 48 processors with the input file \verb|ph.in| and the output file \verb|ph.out|. In step 3, the electron-phonon matrix element $\mathcal{M}_{\text{ep}}^{i \to f}(\mathbf{k},\mathbf{q}\nu)$ (see Eq.~\eqref{eq:elph1}) is calculated using the calculated $\psi_\mathbf{k}$ in the step 1.
    \item[-] \textbf{Step 4}: Run \verb|raman.x| of {\sc QERaman} with a single processor with the input file \verb|raman.in| and the output file \verb|raman.out|. In step 4, the complex Raman tensor $\mathbf{R}(\nu)$ (see Eq.~\eqref{eq:5}) and the Raman spectra $I(E_L, E_{RS})$ (see Eq.~\eqref{eq:4}) are calculated using the calculated $\mathcal{D}^{i \to f}(\mathbf{k})$ and $\mathcal{M}_{\text{ep}}^{i \to f}(\mathbf{k},\mathbf{q}\nu)$ in the steps 2 and 3, respectively. It is noted that \verb|raman.x| supports only the serial processor.
\end{itemize}

\section{Examples}
\label{sec:example}
In this section, we show how to run {\sc QERaman} for 4.1 graphene and 4.2 monolayer MoS$_2$ by showing the input and output files that are stored in \verb|example/graphene| and \verb|example/mos2|, respectively. Before practicing with these tutorials, please ensure that {\sc QE} and {\sc QERaman} have been installed on your computer (see Sec.~\ref{sec:install}). The source files and scripts of these tutorials are also available online on GitHub (\href{https://github.com/nguyen-group/QERaman/tree/main/examples}{https://github.com/nguyen-group/QERaman/tree/main/examples}). The readers can download all input files or read them at GitHub without typing.

\subsection{Graphene}
\label{sec:graphene}
\noindent\ding{111}\enspace\textbf{Purpose:} to calculate the helicity-dependent Raman spectra of graphene for a given laser energy and for the incident/scattered circularly polarized light.

\noindent\ding{111}\enspace\textbf{How to run:} There are four steps to run this tutorial, as shown in Sec.~\ref{sec:run}.

%\noindent\ding{111}\enspace\textbf{How to check:} To check whether or not the output file exists, the readers can use the \verb|ls| command by typing:

\noindent\ding{111}\enspace\textbf{Input files:} Now, let us explain the detail of the input file of each step.

\begin{figure}[t]
  \centering \includegraphics[width=7cm]{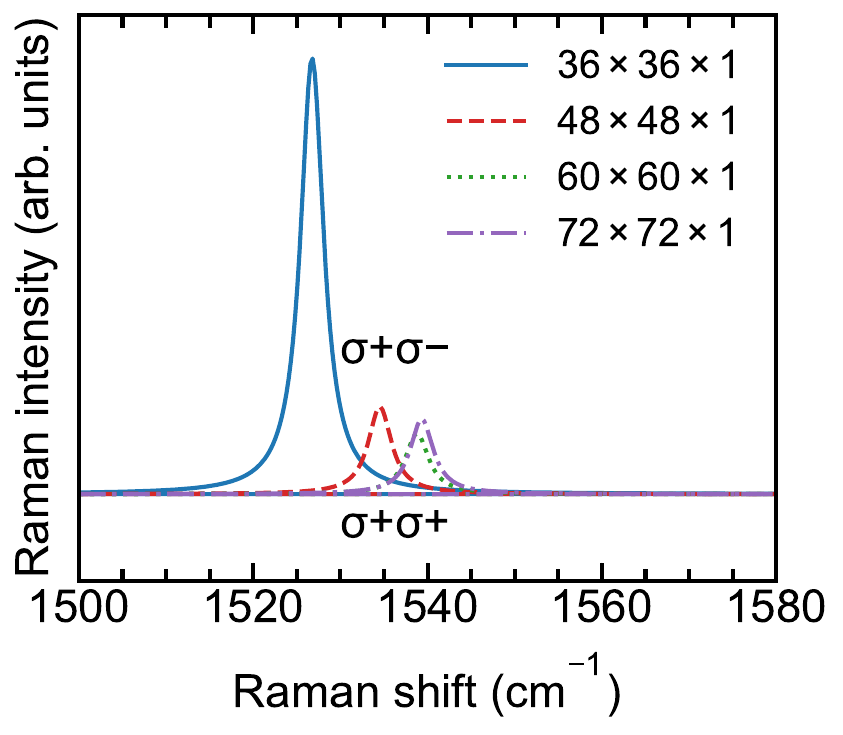}
  \caption{\label{fig:grraman} Helicity-dependent Raman spectra of graphene at the laser energy of 2.33 eV for several k-points grids.}
\end{figure}

For step 1, the input file \verb|scf.in| is given as follows:
\begin{codeQEsmall}{examples/graphene/scf.in}
&CONTROL
  calculation     = 'scf'
  prefix          = 'graphene'
  verbosity       = 'high'
  tstress         = .true.
  tprnfor         = .true.
  outdir          = './tmp'
  pseudo_dir      = './'
/
&SYSTEM
  ibrav           = 4
  celldm(1)       = 4.6608373919
  celldm(3)       = 8.1089553887
  nat             = 2
  ntyp            = 1
  nbnd            = 8
  ecutwfc         = 120.0
  occupations     = 'smearing'
  smearing        = 'mv'
  degauss         = 0.01
  assume_isolated = '2D'
  nosym           = .true.
  noinv           = .true.
/
&ELECTRONS
  conv_thr        = 10.D-10
/
ATOMIC_SPECIES
  C  12.0107  C.pz-hgh.UPF
ATOMIC_POSITIONS (crystal)
  C  0.3333333333  0.6666666667  0.5000000000
  C  0.6666666667  0.3333333333  0.5000000000
K_POINTS (automatic)
  60  60  1  0  0  0
\end{codeQEsmall}
The general explanation of \verb|scf.in| is given in the hands-on guidebook of {\sc QE}~\cite{hung2022quantum} or on the web page: \href{https://www.quantum-espresso.org/Doc/INPUT\_PW.html}{https://www.quantum-espresso.org/Doc/INPUT\_PW.html}. It is noted that we use \verb|nosym = .true.| and \verb|noinv = .true.| at lines 22 and 23, respectively, to unset the crystal symmetry (i.e., k points are expanded to cover the entire Brillouin zone) and time-reversal symmetry (i.e., disable the usage of equivalent k and -k points). The k-points grid $60\times 60 \times 1$ at line 34 is used based on convergence, as shown in Fig.~\ref{fig:grraman}. If the calculation normally finishes, a message \verb|JOB DONE| is written at the end of the output file \verb|scf.out|.

\begin{table*}[t]
\centering
 \begin{tabular}{p{0.4cm} p{2.8cm} p{13.8cm}} 
 \toprule
 \textbf{Line} &\textbf{Syntax} & \textbf{Meaning}\\ \midrule
 1& \texttt{\&inputraman} & A namelist includes input variables for the Raman calculation.\\ \hdashline
 2& \texttt{prefix} & Filenames of output data of {\sc QE}.\\ \hdashline
 3& \texttt{outdir} & Directory name of output files.\\ \hdashline
 4& \texttt{fil\_dvec} & File name of dipole vectors calculated by \texttt{bands\_mat.x}.\\ \hdashline
 5& \texttt{fil\_elph} & File name of electron-phonon matrix elements calculated by \texttt{ph\_mat.x}.\\ \hdashline
 6& \texttt{sorb} & \texttt{.true.} is set if SCF calculation considers spin-orbit interaction. The default is \texttt{.false.}\\ \hdashline
 7& \texttt{circular\_pol} &  \texttt{.true.} is set for calculating Raman spectra for circularly-polarized light. The default is \texttt{.false.}\\ \hdashline
 8& \texttt{nonpol} & \texttt{.true.} is set for calculating Raman spectra for non-polarized light. The default is \texttt{.false.}\\ \hdashline
 9& \texttt{plot\_matele\_opt} & \texttt{.true.} is set for plotting the dipole vectors. The default is \texttt{.false.}\\ \hdashline
 10& \texttt{plot\_matele\_elph} & \texttt{.true.} is set for plotting the electron-phonon matrix elements. The default is \texttt{.false.}\\ \hdashline
 11& \texttt{plot\_raman\_k} & \texttt{.true.} is set for plotting the Raman matrix element. The default is \texttt{.false.}\\ \hdashline
 12& \texttt{gamma} & Broadening factor of resonance condition $\gamma$ in the eV unit. The default value is 0.1 eV\\ \hdashline
 13& \texttt{gamma\_raman} & Broadening factor of Raman spectra $\Gamma$ in the eV unit. The default value is 0.00005 eV\\ \hdashline
 14& \texttt{rs\_start} & Starting value of calculated Raman shift in the cm$^{-1}$ unit. The default value is 0.0 cm$^{-1}$.\\ \hdashline
 15& \texttt{rs\_end} & Ending value of calculated Raman shift in the cm$^{-1}$ unit. The default value is 0.0 cm$^{-1}$.\\ \hdashline
 16& \texttt{nrs} & Number of calculated points of Raman shift. The default value is 500.\\ \hdashline
 17& \texttt{elaser1} & Laser energies in the eV unit. The default value is 0.1 eV. We can calculate up to 7 laser energy in one calculation, i.e., \texttt{elaser1}, $\ldots$, \texttt{elaser7}.\\ \hdashline
 18& \texttt{/} & End of namelist \texttt{\&inputraman}.\\
 \bottomrule
 \end{tabular}
\caption{Meaning of input variables in \texttt{raman.in} file.}
\label{tab:raman}
\end{table*}

For step 2, the input file \verb|bands.in| is given as follows:
\begin{codeQEsmall}{examples/graphene/bands.in}
&BANDS
  prefix   = 'graphene'
  outdir   = './tmp'
  filband  = 'graphene.bands'
  lp       = .true.
  filp     = 'graphene.dvec'
/
\end{codeQEsmall}
The commands from lines 1 to 6 are explained in detail on the web page: \href{https://www.quantum-espresso.org/Doc/INPUT\_BANDS.html}{https://www.quantum-espresso.org/Doc/INPUT\_BANDS.html}. It is noted that the syntax \verb|lp = .true.| in the \verb|bands_mat.x| command is used to obtain the complex number of dipole vectors (see \ref{sec:elpt}). 
It is important to note that \verb|lp = .true.| is used in \verb|bands.x| for the original {\sc QE}. However, the original {\sc QE} gives the absolute value of the electron-photon matrix elements. Thus the users use the modified \verb|bands.x| in this folder. We intentionally use the different name of \verb|bands_mat.x| to distinguish from \verb|bands.x|.
If the calculation normally finishes, a message \verb|JOB DONE| is written at the end of the output file \verb|bands.out|, and the dipole vectors are written in the output file \verb|graphene.dvec|.

For step 3, the input file \verb|ph.in| is given as follows:
\begin{codeQEsmall}{examples/graphene/ph.in}
phonons
&inputph
  prefix          = 'graphene'
  outdir          = './tmp'
  tr2_ph          = 1.0d-18
  verbosity       = 'high'
  fildyn          = 'graphene.dyn'
  fildvscf        = 'dvscf'
  electron_phonon = 'epc'
/
  0.0  0.0  0.0
\end{codeQEsmall}
The commands from lines 1 to 11 are explained in detail on the web page: \href{https://www.quantum-espresso.org/Doc/INPUT\_PH.html}{https://www.quantum-espresso.org/Doc/INPUT\_PH.html}. It is noted that the syntax \verb|electron_phonon = 'epc'| in the \verb|ph_mat.x| command is used to obtain the complex number of electron-phonon matrix elements (see \ref{sec:elph}). Thus, the present input file can not be used in \verb|ph.x| in the original {\sc QE}. If the calculation normally finishes, a message \verb|JOB DONE| is written at the end of the output file \verb|ph.out|, and the electron-phonon matrix elements are written in the output file \verb|graphene.elph|.

For step 4, the input file \verb|raman.in| is given as follows:
\begin{codeQEsmall}{examples/graphene/raman.in}
&inputraman
  prefix           = 'graphene'
  outdir           = './'
  fil_dvec         = './graphene.dvec'
  fil_elph         = './graphene.elph'
  sorb             = .false.
  circular_pol     = .true.
  nonpol           = .false.
  plot_matele_opt  = .false.
  plot_matele_elph = .false.
  plot_raman_k     = .false.
  gamma            = 0.05
  gamma_raman      = 0.0002
  rs_start         = 1500.0
  rs_end           = 1600.0
  nrs              = 400
  elaser1          = 2.33
/
\end{codeQEsmall}
The commands from lines 1 to 20 are explained in detail in Table~\ref{tab:raman}. If the calculation normally finishes, a message \verb|JOB DONE| is written at the end of the output file \verb|raman.out|, and the Raman spectra are written in the output file \verb|raman_spectra1.dat|.

\noindent\ding{111}\enspace\textbf{Output files:} First, we will show the helicity-dependent Raman spectra for the circularly polarized light using Eq.~\eqref{eq:1}, as shown in Fig.~\ref{fig:grraman}. The calculated Raman spectra are obtained by the \verb|raman_spectra1.dat| file for the laser energy of 2.33 eV for $36\times 36 \times 1$, $48\times 48 \times 1$, $60\times 60 \times 1$, and $72\times 72 \times 1$ k-points grids. A Jupyter notebook file \verb|plot-raman-spectra.ipynb| in \verb|example/graphene/reference| is used to plot the Raman spectra in Fig.~\ref{fig:grraman}. Due to the resonance condition $[E_L-\Delta E_{mi}(\mathbf{k})][E_L-\Delta E_{m^{\prime}i}(\mathbf{k})-\hbar\omega_\nu]$ in Eq.~\eqref{eq:1}, both Raman intensity and peak position are sensitive to the selection of the k-points grids. 
%Thus, to compare the Raman spectra for different k-points grids, the Raman spectra are normalized by the maximum value of the Raman peak. 
The calculated results show that the $60\times 60 \times 1$ is a suitable k-points grid for the convergence of the Raman peak. In the case of semimetal graphene, the Raman scattering occurs near the Dirac point. Thus, a dense k-point grid is required to obtain the convergence results. 

The Raman spectra of graphene also show the nonzero Raman intensity at the $G$ band (1540 cm$^{-1}$) for helicity-changing ($\sigma + \sigma -$) scattering, while the zero Raman intensity for helicity-conserving ($\sigma + \sigma +$) scattering, which is consistent with the experiment by Drapcho \textit{et al.}~\cite{drapcho2017apparent}. This result is also consistent with the conservation of angular momentum, which implies that the doubly-degenerate phonon mode ($E_{2g}$) can change the helicity of circularly-polarized light~\cite{tatsumi2018conservation}.
%The helicity-exchange phenomenon of graphene can be also obtained by using the group theory. Graphene belongs to the $D_{6h}$ point group, as shown in Table 3. The polarization vectors thus belong to the irreducible representation $\Gamma^{-}_5$ with the two basis functions $|5,1\rangle^{-}$ and $|5,2\rangle^{-}$, which correspond to the left- and right-handed circular polarization. 

Next, we will show the polarized Raman intensity with the linearly-polarized light from the calculated, complex Raman tensor. Here, we consider that linearly-polarized light propagates in the direction perpendicular to the graphene plane and that the polarization vectors for the incident and scattered light are parallel to each other, which we call VV configuration. In this case, the Raman intensity is given by~\cite{saito2016raman}
\begin{equation}
\label{eq:10}
I(\nu,\theta)=
  \left|\,\begin{pmatrix}\, \cos{\theta} & \sin{\theta} & 0 \,\end{pmatrix}\mathbf{R}(\nu)\begin{pmatrix} \,\cos{\theta} \\ \sin{\theta} \\ 0\, \end{pmatrix}\,\right|^2.
\end{equation}

\begin{figure}[t]
  \centering \includegraphics[width=8cm]{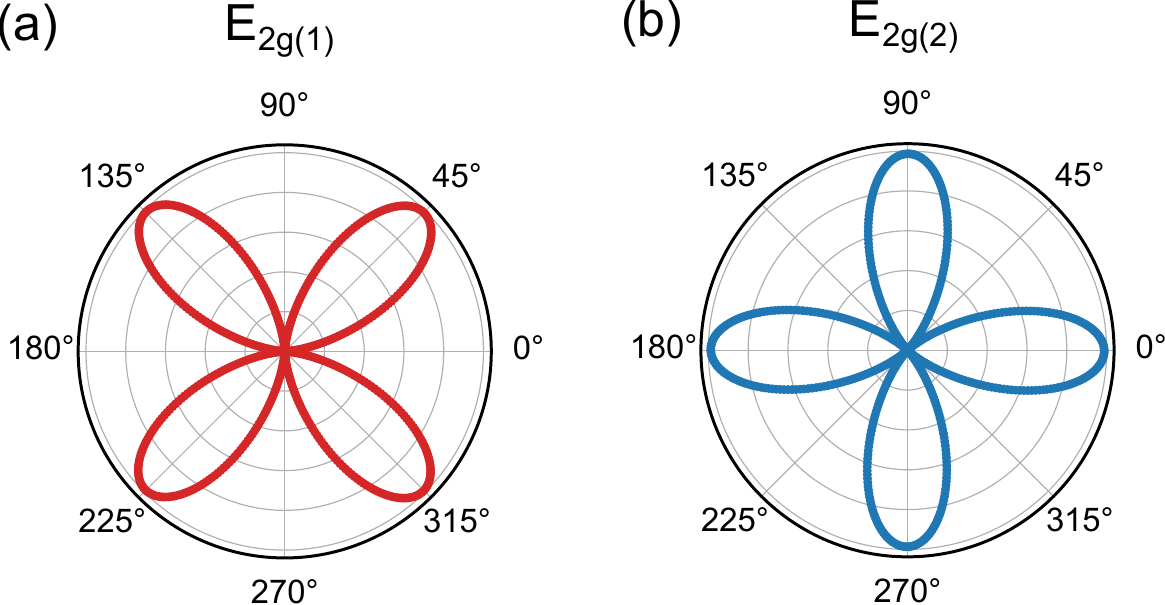}
  \caption{\label{fig:grraman-polar} Polarized Raman intensities for linearly-polarized light are plotted for the doubly-degenerate $E_{2g(1)}$ (a) and $E_{2g(2)}$ (b) based on the complex Raman tensors.}
\end{figure}

Graphene has two carbon atoms per unit cell, leading to six phonon modes, including three acoustic and three optical phonon modes, and the symmetry of graphene belongs to the $D_{6h}$ point group. Therefore, the irreducible representations of phonon modes at the $\Gamma$ point are given by~\cite{saito1998physical}
\begin{equation}
\label{eq:7}
\Gamma^{\text{graphene}}=E_{1u}+A_{2u}+B_{1g}+E_{2g},
\end{equation}
where $E_{1u}$ represents the translation in the graphene plane (iTA+LA phonon), $A_{2u}$ represents translation perpendicular to the graphene plane (or oTA phonon), $B_{2g}$ is the Raman-inactive optical oTO phonon (the carbon atoms move perpendicularly to the graphene plane), and $E_{2g}$ is the doubly-degenerate Raman active optical iTO + LO phonon (the carbon atoms move in the graphene plane). The complex Raman tensors of $E_{2g(1)}$ and $E_{2g(2)}$ are given by
\begin{equation}
\label{eq:8}
\mathbf{R}(E_{2g(1)},xy)=
  \begin{pmatrix}
    0      & {a+ib} & 0 \,\\
    {a+ib} & 0      & 0 \,\\
    0      & 0      & 0 \,
  \end{pmatrix},
\end{equation}
and
\begin{equation}
\label{eq:9}
\mathbf{R}(E_{2g(2)},x^2-y^2)=
  \begin{pmatrix}
    {a+ib} & 0       & 0 \,\\
    0      & {-a-ib} & 0 \,\\
    0      & 0       & 0 \,
  \end{pmatrix},
\end{equation}
where $a$ and $b$ are real and imaginary part of complex Raman tensor, respectively. 
The calculated complex Raman tensors are written at the end of the \verb|raman.out| file, 
which shows that $a=0.10$ and $b=0.99$.
By inserting Eqs.~\eqref{eq:8} and~\eqref{eq:9} into Eq.~\eqref{eq:10}, the polarized Raman intensities for $E_{2g(1)}$ and $E_{2g(2)}$ are given as follows
\begin{equation}
\label{eq:11}
I(E_{2g(1)},\theta)=4(a^2+b^2)\cos^2{\theta}\sin^2{\theta}
\end{equation}
and
\begin{equation}
\label{eq:12}
I(E_{2g(2)},\theta)=(a^2+b^2)\cos^2{2\theta},
\end{equation}
respectively.  After obtaining the complex Raman tensor, the reader can use a Jupyter notebook file \verb|plot-raman-polar.ipynb| in \verb|example/graphene/reference| to plot the polarized Raman intensity. In Figs.~\ref{fig:grraman-polar}(a) and (b), we show the Raman intensity for $E_{2g(1)}$ and $E_{2g(2)}$ as a function of polarization angle $\theta$, respectively. The Raman intensity for $E_{2g(1)}$ ($E_{2g(2)}$) has maximum intensity $I_{\text{max}}=a^2+b^2$ for $\theta=45^{\circ}, 135^{\circ}, 225^{\circ}$, and $315^{\circ}$ ($\theta=0^{\circ}, 90^{\circ}, 180^{\circ}$, and $270^{\circ}$). The Raman tensor $a$ and $b$ values depend on the input parameters \verb|gamma| and \verb|elaser1|, as shown in Eq.~\eqref{eq:5}. Thus, the values of \verb|gamma| and \verb|elaser1| are essential input parameters and need to compare with the experimental data. 

\subsection{MoS\texorpdfstring{$_2$}{} monolayer}
\label{sec:mos2}
\noindent\ding{111}\enspace\textbf{Purpose:} to calculate the helicity-dependent Raman spectra as a function of the laser energy for the circularly polarized light for monolayer MoS$_2$.

\noindent\ding{111}\enspace\textbf{How to run:} There are four steps to run this tutorial, as shown in Sec.~\ref{sec:run}.

%\noindent\ding{111}\enspace\textbf{How to check:} To check whether or not the output file exists, the readers can use the \verb|ls| command by typing:

\noindent\ding{111}\enspace\textbf{Input files:} All input files are given as follows:
\begin{codeQEsmall}{examples/mos2/scf.in}
&CONTROL
  calculation     = 'scf'
  prefix          = 'MoS2'
  verbosity       = 'high'
  tstress         = .true.
  tprnfor         = .true.
  outdir          = './tmp'
  pseudo_dir      = './'
/
&SYSTEM
  ibrav           = 4
  celldm(1)       = 5.9329994115
  celldm(3)       = 7.9627773219
  nat             = 3
  ntyp            = 2
  nbnd            = 15
  ecutwfc         = 120.0
  occupations     = 'fixed'
  assume_isolated = '2D'
  nosym           = .true.
  noinv           = .true.
/
&ELECTRONS
  conv_thr        = 1.0D-10
/
ATOMIC_SPECIES
  Mo  95.942  Mo.pz-hgh.UPF
  S   32.065  S.pz-hgh.UPF
ATOMIC_POSITIONS (crystal)
  Mo  0.3333333333  0.6666666667  0.5000000000
  S   0.6666666667  0.3333333333  0.5624768914
  S   0.6666666667  0.3333333333  0.4375231086
K_POINTS (automatic)
  48  48  1  0  0  0
\end{codeQEsmall}

\begin{codeQEsmall}{examples/mos2/bands.in}
&BANDS
  prefix   = 'MoS2'
  outdir   = './tmp'
  filband  = 'MoS2.bands'
  lp       = .true.
  filp     = 'MoS2.dvec'
/
\end{codeQEsmall}

\begin{codeQEsmall}{examples/mos2/ph.in}
phonons
&inputph
  prefix          = 'MoS2'
  outdir          = './tmp'
  tr2_ph          = 1.0D-18
  verbosity       = 'high'
  fildyn          = 'MoS2.dyn'
  fildvscf        = 'dvscf'
  electron_phonon = 'epc'
/
  0.0  0.0  0.0
\end{codeQEsmall}

\begin{codeQEsmall}{examples/mos2/raman.in}
&inputraman
  prefix           = 'MoS2'
  outdir           = './'
  fil_dvec         = './MoS2.dvec'
  fil_elph         = './MoS2.elph'
  sorb             = .false.
  circular_pol     = .true.
  nonpol           = .false.
  plot_matele_opt  = .false.
  plot_matele_elph = .false.
  plot_raman_k     = .false.
  gamma            = 0.5
  gamma_raman      = 0.0002
  rs_start         = 350.0
  rs_end           = 450.0
  nrs              = 400
  elaser1          = 1.95
  elaser2          = 2.33
  elaser3          = 2.54
  elaser4          = 2.78
/
\end{codeQEsmall}
The detail of the input parameters is similar to the input parameters of the example with graphene (see Sec.4.1). It is noted that the k-point grid of $48\times 48 \times 1$ is selected by the convergence test.

\begin{figure}[t]
  \centering \includegraphics[width=6cm]{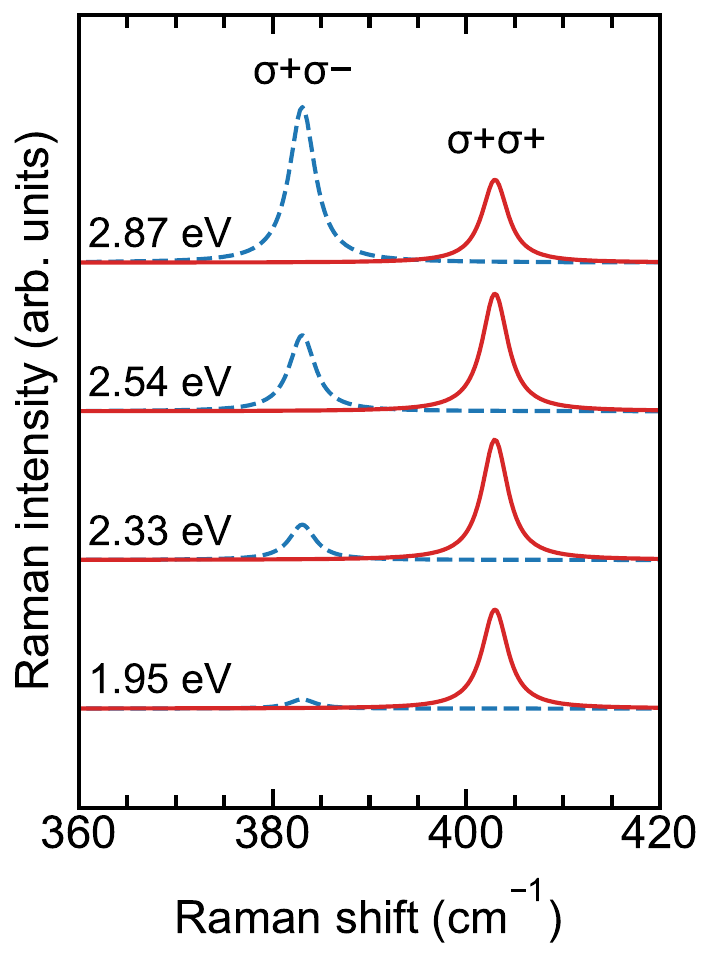}
  \caption{\label{fig:mos2raman} Helicity-dependent Raman spectra of monolayer MoS$_2$ at 1.95, 2.33, 2.54, and 2.87 laser energies.}
\end{figure}

\noindent\ding{111}\enspace\textbf{Output files:} First, we plot the helicity-dependent Raman spectra for the circularly-polarized light by using \verb|plot-raman-spectra.ipynb| in \verb|example/mos2/reference|. The Raman spectra are obtained by the \verb|raman_spectra1.dat|, \verb|raman_spectra2.dat|, \verb|raman_spectra3.dat|, and \verb|raman_spectra4.dat| files for 1.95, 2.33, 2.54, and 2.87 eV laser energies, respectively. As shown in Fig.~\ref{fig:mos2raman}, the Raman peaks are found at 383 cm$^{-1}$ (in-plane metal-and-chalcogen (IMC) phonon mode) and 403 cm$^{-1}$ (out-of-plane chalcogen (OC) phonon), which are in agreement with the observed Raman (384 cm$^{-1}$ and 403 cm$^{-1}$, respectively~\cite{lee2010anomalous,li2012quantitative,sun2013spin}). The IMC mode is the in-plane vibration of the Mo and S atoms, while the OC mode is the out-of-plane vibration of S atoms. The IMC peak shows nonzero Raman intensities for helicity-changing $\sigma+\sigma-$, while the OC peak shows the nonzero Raman intensities for helicity-conserving $\sigma+\sigma+$, which is consistent with the experimental data by Chen \textit{et al.}~\cite{chen2015helicity}.

As shown in Fig.~\ref{fig:mos2raman}, the relative intensity of the IMC to the OC depends on the incident laser energy $E_L$, in which the IMC peak increases with increasing $E_L$. Tatsumi \textit{et al.}~\cite{tatsumi2016laser} showed that the electron-photon matrix elements $\mathcal{M}_{\text{opt}}(\mathbf{k})$ depends on $E_L$, in which $\mathcal{M}_{\text{opt}}(\mathbf{k})$ becomes larger with increasing $E_L$, and the optical transition not only occurs around the K and K$^{\prime}$ points but also the $\Lambda$ and $\Lambda^{\prime}$ points. Thus the Raman intensity of the IMC mode is enhanced by increasing $E_L$, while the Raman intensity of the OC mode is suppressed at $E_L=2.87$ eV because of the contribution of the smaller $\mathcal{M}_{\text{ep}}(\mathbf{k},\mathbf{q}\nu)$ around the M point compared with $\mathcal{M}_{\text{ep}}(\mathbf{k},\mathbf{q}\nu)$ around the $\Lambda(\Lambda^{\prime})$ point~\cite{tatsumi2016laser} at $E_L=2.87$ eV. We note that since the $\sigma+(\sigma-)$ light is absorbed only at the K(K$^{\prime}$) point, the helicity-changing IMC peak is not suppressed by the $\mathcal{M}_{\text{ep}}(\mathbf{k},\mathbf{q}\nu)$ around M point. It notes that the Raman intensity of the IMC peak is comparable to that of the OC peak in the experiment for both 2.33 eV and 2.45 eV~\cite{chen2015helicity}, while the calculated result shows the IMC peak is weaker than the OC peak at 2.33 eV. One possible reason for the disagreement is the evaluation of the input parameter \verb|gamma| (or $\gamma$ in Eq.~\eqref{eq:1}, which is related to the lifetime of the photo-excited carrier). The $\gamma$ value should differ for the IMC and OC modes. In particular, $\gamma$ of the IMC mode should be longer than that of the OC mode since the photoexcited carrier in the IMC Raman process cannot relax to the ground state by valley polarization. On the other hand, $\gamma$ should change by changing $E_L$ due to the optical transition that occurs from the K (K$^{\prime}$) points to the $\Lambda$ ($\Lambda^{\prime}$) points when increasing $E_L$~\cite{tatsumi2016laser}. Since the $\gamma$-dependent phonon mode and laser energy is evaluated by
calculating the whole path of the electron-photon and electron-
phonon scattering, it is beyond the present code. Another possible reason is the exciton effect on the monolayer MoS$_2$. The present code supports the standard QE calculation for the electronic band structure (i.e., using LDA or GGA), which is known to underestimate the band gap~\cite{hung2022quantum}. This problem can be treated by shifting $E_L$ by a different band-gap value between the QE calculation with the experiment data. It is noted that this treatment is a simple approximation since the exciton effect modifies not only the electronic band gap but also the electronic wavefunctions.

\begin{figure}[t]
  \centering \includegraphics[width=8cm]{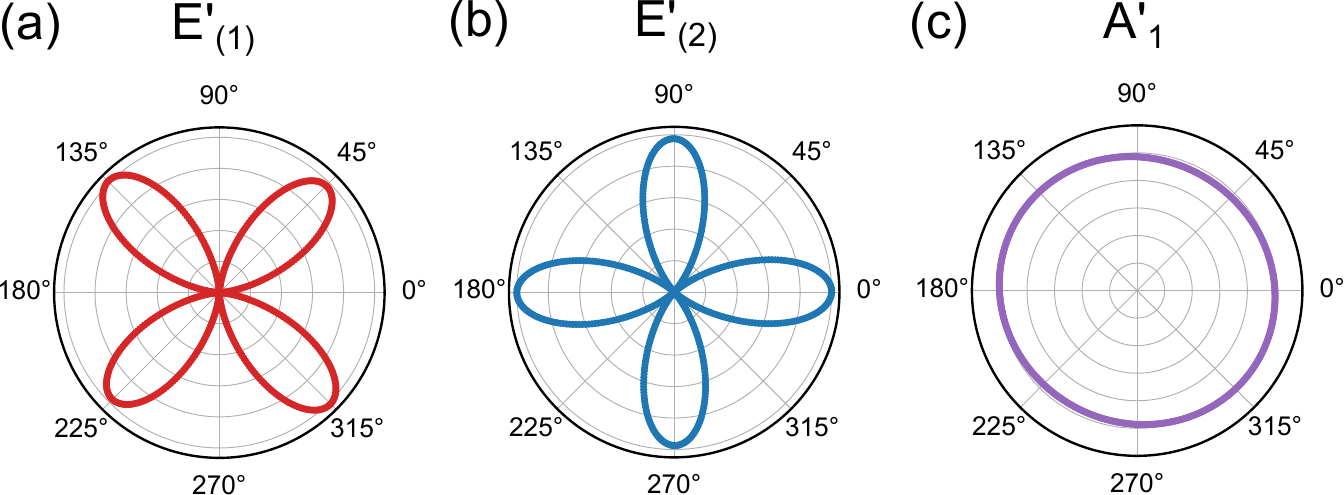}
  \caption{\label{fig:mos2raman-polar} Polarized Raman intensities are plotted for $E_{2g(1)}$ (a), $E_{2g(2)}$ (b), and $A_{1g}$ (c) based on the complex Raman tensors at 2.54 eV.}
\end{figure}

Next, we plot the polarized Raman intensities for the Raman activity modes with the linearly-polarized light by using \verb|plot-raman-polar.ipynb| in \verb|example/mos2/reference|. The monolayer MoS$_2$ belongs to D$_{3h}$ point group with six optical and three acoustic phonon modes. Thus, the irreducible representations at the $\Gamma$ point are given by~\cite{saito2016raman}
\begin{equation}
\label{eq:13}
\Gamma^{\text{MoS2}}=2A^{\prime\prime}_2+2E^{\prime}+E^{\prime\prime}+A^{\prime}_1,
\end{equation}
where the doubly-degenerate $E^{\prime}$ (IMC mode) and $A^{\prime}_1$ (OC mode) have the Raman activity. The complex Raman tensors of these modes can be found in the \verb|raman.out| file. By using Eq.~\eqref{eq:10}, we can plot the Raman intensity as a function of the polarization angle $\theta$ for the doubly-degenerate $E^{\prime}$ ($E^{\prime}_{(1)}$ and $E^{\prime}_{(2)}$) modes and the $A^{\prime}_1$ mode, as shown in Fig.~\ref{fig:mos2raman-polar}(a)-(c). The shapes of the $E^{\prime}$ mode are similar to that of the $E_{2g}$ of the graphene, while the $A^{\prime}_1$ mode does not depend on $\theta$.

\section{Summary}
\label{sec:summary}
In summary, we developed a new open-source code {\sc QERaman} for computing the first-order resonance Raman and the complex Raman tensors based on the output of {\sc Quantum ESPRESSO}, which is free and widely-used software. We explain how to download and install the program. We also show how to use the program through hands-on tutorials with graphene and the monolayer MoS$_2$. Since our program is designed as an open-source module of the {\sc Quantum ESPRESSO} distribution, {\sc QERaman} should be user-friendly for the Quantum ESPRESSO community. The program is aimed at experimentalists who want a theoretical tool for understanding the observed resonance Raman spectra for the linearly- or circularly-polarized light. The authors would like to know any comments from users. 

%\clearpage
%\newpage

% Appendix
\appendix

\section{Electron-photon matrix elements}
\label{sec:elpt}
In this section, we discuss the electron-photon matrix elements and the dipole vector, which can be obtained from the \verb|bands_mat.x| in the {\sc QERaman} code. Let us consider an electron of mass $m$ and charge $e$ in an electromagnetic field. The Hamiltonian of the electron is given by:
\begin{equation}
    \label{eq:elpt1}
    \mathcal{H}=\frac{1}{2m}(-i\hbar\nabla + e\mathbf{A})^2 + V(\mathbf{r}) - e\Phi,
\end{equation}
where $\mathbf{A}$ is the vector potential, and $\phi$ is the scalar potential. From Eq.~\eqref{eq:elpt1}, the Schr\"{o}dinger equation is written as
\begin{strip}
\begin{equation}
    \label{eq:elpt2}
    \begin{split}
        E\psi&=\left[\frac{1}{2m}(-i\hbar\nabla + e\mathbf{A})^2 + V(\mathbf{r})- e\Phi\right]\psi =\frac{1}{2m}(-i\hbar\nabla + e\mathbf{A})\cdot(-i\hbar\nabla\psi + e\mathbf{A}\psi) +  V(\mathbf{r})\psi- e\Phi\psi\\
        &=-\frac{\hbar^2}{2m}\nabla^2\psi - \frac{ie\hbar}{2m}\nabla\cdot(\mathbf{A}\psi)- \frac{ie\hbar}{2m}\mathbf{A}\cdot\nabla\psi+\frac{e^2}{2m}\mathbf{A}^2\psi + V(\mathbf{r})\psi- e\Phi\psi\\
        &=-\frac{\hbar^2}{2m}\nabla^2\psi-\frac{ie\hbar}{2m}(\nabla\cdot\mathbf{A})\psi- \frac{ie\hbar}{m}\mathbf{A}\cdot\nabla\psi+\frac{e^2}{2m}\mathbf{A}^2\psi + V(\mathbf{r})\psi- e\Phi\psi,
    \end{split}
\end{equation}
\end{strip}
where we used that $\nabla\cdot(\mathbf{A}\psi) = (\nabla\cdot\mathbf{A})\psi+\mathbf{A}\cdot\nabla\psi$.
By adopting the Coulomb gauge for $\mathbf{A}$
\begin{equation}
    \label{eq:elpt2a}
    \nabla\cdot\mathbf{A}=0, \Phi =0,
\end{equation}
and neglecting the second-order term of $\mathbf{A}$, Eq.~\eqref{eq:elpt2} is rewritten as
\begin{equation}
    \label{eq:elpt3}
    E\psi=\left[ -\frac{\hbar^2}{2m}\nabla^2 + V(\mathbf{r}) -  \frac{ie\hbar}{m}\mathbf{A}\cdot\nabla\right]\psi.
\end{equation}
Then the Hamiltonian in Eq.~\eqref{eq:elpt1} is rewritten as
\begin{equation}
    \label{eq:elpt4}
    \mathcal{H}={\underbrace{-\frac{\hbar^2}{2m}\nabla^2 + V(\mathbf{r})}_{\mathcal{H}_0}} +{\underbrace{  -\frac{ie\hbar}{m}\mathbf{A}\cdot\nabla}_{\mathcal{H}_{\text{opt}}}},
\end{equation}
where $\mathcal{H}_0$ and $\mathcal{H}_{\text{opt}}$ are the unperturbed Hamiltonian and perturbation Hamiltonian for electron-photon interaction, respectively. By given $\mathcal{H}_{\text{opt}}$ in Eq.~\eqref{eq:elpt4}, the electron-photon matrix element $\mathcal{M}_{\text{opt}}^{i\to f}$ for a pair of initial state $i$ and final state $f$ is defined as
\begin{equation}
    \label{eq:elpt5}
    \mathcal{M}_{\text{opt}}^{i \to f}(\mathbf{k})=\langle \psi^{f}_\mathbf{k}|\mathcal{H}_{\text{opt}}| \psi^{i}_\mathbf{k}\rangle=-\frac{ie\hbar}{m}\langle \psi^{f}_\mathbf{k}|\mathbf{A}\cdot\nabla| \psi^{i}_\mathbf{k} \rangle,
\end{equation}
where $\psi^{f}_\mathbf{k}$ and $\psi^{i}_\mathbf{k}$ are the Kohn-Sham orbitals of the $f$ and $i$ states at the wavevector $\mathbf{k}$.
 
Here, we consider an electromagnetic field with no electric charges and currents. In this case, Maxwell's equation reduces to~\cite{kittel1976introduction}
\begin{equation}
    \label{eq:elpt6}
    \nabla\times \mathbf{B}=\frac{1}{c}\frac{\partial \mathbf{E}}{\partial t},
\end{equation}
where $c$ is the speed of light in the vacuum, and the electric field $\mathbf{E}$ and the magnetic field $\mathbf{B}$ are given by
\begin{equation}
    \label{eq:elpt7}
  \begin{cases}
    \mathbf{E}=\displaystyle{-\frac{1}{c}\frac{\partial \mathbf{A}}{\partial t}}-\nabla\Phi\\
    \mathbf{B}=\nabla\times\mathbf{A}
  \end{cases}
\end{equation}
By inserting Eq.~\eqref{eq:elpt7} into Eq.~\eqref{eq:elpt6}, we obtain the following equation,
\begin{equation}
    \label{eq:elpt8a}    \nabla\times(\nabla\times\mathbf{A})=\nabla(\nabla\cdot\mathbf{A})-\nabla^2 \mathbf{A}=-\frac{1}{c^2}\frac{\partial^2\mathbf{A}}{\partial t^2}-\nabla\left(\frac{1}{c}\frac{\partial\Phi}{\partial t}\right).
\end{equation}
Then applying the condition in Eq.~\eqref{eq:elpt2a}, Eq.~\eqref{eq:elpt8a} becomes
\begin{equation}
    \label{eq:elpt8}
    \left(\nabla^2-\frac{1}{c^2}\frac{\partial^2 }{\partial t^2}\right)\mathbf{A}=\nabla\left[\nabla\cdot \mathbf{A} +\frac{1}{c}\frac{\partial}{\partial t}\Phi\right]=0.
\end{equation}
The plane-wave solution of Eq.~\eqref{eq:elpt8} has the following form:
\begin{equation}
    \label{eq:elpt10}
    \mathbf{A}=\left(\mathcal{A} \exp^{i(\mathbf{k}_{\text{opt}}\mathbf{r}-\omega_{\text{opt}} t)}+\text{ c.c.}\right)\mathbf{P},
\end{equation}
where $\mathcal{A}$ is a complex number (the amplitude) that specifies the magnitude and phase of the plane wave, c.c. denotes the complex conjugate of the first term, and $\mathbf{P}$ is a unit vector (or the polarization vector or the Jones vector) that specifies the direction of the vector potential $\mathbf{A}$.  $\omega_{\text{opt}}$ and $\mathbf{k}_{\text{opt}}$ are the angular frequency and wavevector of light, respectively, which satisfies the dispersion of a photon,
\begin{equation}
    \label{eq:elpt11}
    \omega_{\text{opt}}=c|\mathbf{k}_{\text{opt}}|. 
\end{equation}
Further,  inserting Eq.~\eqref{eq:elpt10} into $\nabla\cdot\mathbf{A}=0$ in Eq.~\eqref{eq:elpt2a}, we obtain
\begin{equation}
    \label{eq:elpt12}
    \mathbf{k}_{\text{opt}}\cdot \mathbf{P} = 0. 
\end{equation}
Thus, for a given $\mathbf{k}_{\text{opt}}$, we have two orthogonal polarization vectors $\mathbf{P}$. A general expression of $\mathbf{P}$ for an electromagnetic wave propagating in the $z$ direction is written as
\begin{equation}
    \label{eq:elpt13}
    \mathbf{P} = \frac{1}{\sqrt{P_x^2+P_y^2}}\left(
    \begin{array}{c}
      P_x \\
      P_y e^{i\phi} \\
      0
    \end{array}
  \right),
\end{equation}
where $P_x(P_y)$ and $\phi$ are, respectively, the amplitude of $x(y)$ component defined by a real number and the phase difference between the $x$ and $y$ components of $\mathbf{P}$.
 $\phi=0$ and  $\phi=\pm \pi/2$ corresponds to linearly-polarized and circularly-polarized light, respectively. 
 A general  $\phi$ is elliptically-polarized light.

By inserting the vector potential $\mathbf{A}$ in Eq.~\eqref{eq:elpt10} into Eq.~\eqref{eq:elpt5}, $\mathcal{M}_{\text{opt}}^{i \to f}(\mathbf{k})$ is rewritten as
\begin{equation}
\label{eq:elpt14}
\begin{split} 
    \mathcal{M}_{\text{opt}}^{i \to f}(\mathbf{k})&=\frac{-ie\hbar}{m}\left(\mathcal{A} \exp^{i(\mathbf{k}_{\text{opt}}-\omega_{\text{opt}} t)}+\text{ c.c.}\right)\langle \psi^{f}_\mathbf{k}|\nabla| \psi^{i}_\mathbf{k} \rangle\cdot \mathbf{P}\\
    &=C_{\text{opt}}\mathcal{D}^{i \to f}(\mathbf{k}) \cdot \mathbf{P}.
\end{split}
\end{equation}
Here, we define the potential associated with light, $C_{\text{opt}}$, as follows:
\begin{equation}
    \label{eq:elpt14a}
C_{\text{opt}}=\frac{-ie\hbar}{m}\left(\mathcal{A} \exp^{i(\mathbf{k}_{\text{opt}}-\omega_{\text{opt}} t)}+ \text{ c.c.}\right),
\end{equation}
and we define the dipole vector $\mathcal{D}^{i \to f}(\mathbf{k})$ as follows~\cite{gruneis2003inhomogeneous}:
\begin{equation}
    \label{eq:elpt15}
    \mathcal{D}^{i \to f}(\mathbf{k}) = \langle \psi^{f}_\mathbf{k}|\nabla| \psi^{i}_\mathbf{k} \rangle.
\end{equation}

The absolute value $|\mathcal{D}^{i \to f}(\mathbf{k})|$ between valence and conduction bands can be obtained by using \verb|bands.x| in {\sc QE} with \verb|lp = .true.| in the namelist \verb|&BANDS|. However, as shown in Eq.~\eqref{eq:1} of the Raman intensity, we need a complex number of $\mathcal{D}^{i \to f}(\mathbf{k})$ for any pair of initial state $i$ and final state $f$. Therefore, a Fortran file in the directory \verb|PP/src/write_p_avg.f90| in {\sc QE} is modified to obtain the complex number of $\mathcal{D}^{i \to f}(\mathbf{k})$.

%The modified \verb|write_p_avg.f90| in the directory \verb|qe-7.1/PP/src/| is shown as below:

\section{Electron-phonon matrix elements}
\label{sec:elph}
In this section, we discuss the electron-phonon matrix elements, $\mathcal{M}_\text{ep}^{i \to f}$, which can be obtained from the \verb|ph_mat.x| in the {\sc QERaman} code. Within density-functional perturbation theory~\cite{baroni1987green,baroni2001phonons}, $\mathcal{M}_\text{ep}^{i \to f}$ can be obtained from the first-order derivative of the self-consistent Kohn-Sham potential~\cite{kohn1965self}, $\mathcal{V}_\text{KS}$, with respect to atomic displacements $u_{s\mathbf{R}}$ for the $s$-th atom in lattice position $\mathbf{R}$ as follows:
\begin{equation}
    \label{eq:elph1}
    \mathcal{M}_{\text{ep}}^{i \to f}(\mathbf{k},\mathbf{q}\nu) = \left(\frac{\hbar}{2\omega_{\mathbf{q}\nu}}\right)^{1/2}\langle \psi_{\mathbf{k}+\mathbf{q}}^f|\Delta \mathcal{V}_\text{KS}^{\mathbf{q}\nu}| \psi_\mathbf{k}^i \rangle,
\end{equation}
where $\omega_{\mathbf{q}\nu}$ is the phonon frequency of the phonon mode $\nu$ at the wavevector $\mathbf{q}$, and $\mathcal{V}_\text{K.S.}^{\mathbf{q}\nu}$ is the self-consistent first order variation of the Kohn-Sham potential, which is given by~\cite{wierzbowska2005origins}
\begin{equation}
    \label{eq:elph2}
    \mathcal{V}_\text{KS}^{\mathbf{q}\nu}=\sum_{\mathbf{R}}\sum_s\frac{\partial \mathcal{V}_\text{KS}}{\partial u_{s\mathbf{R}}} \cdot u_s^{\mathbf{q}\nu}\frac{e^{i\mathbf{q}\mathbf{R}}}{\sqrt{N}},
\end{equation}
where $N$ is the number of cells in the crystal, and $u_s^{\mathbf{q}\nu}$ is the displacement pattern for the phonon mode $\nu$ at the wavevector $\mathbf{q}$. In {\sc QE}, the electron-phonon interaction is controlled by syntax \verb|electron_phonon| in the namelist \verb|&INPUTPH|. By setting \verb|electron_phonon = 'epa'|, the matrix elements $\mathcal{M}_{\text{ep}}^{i \to f}(\mathbf{k},\mathbf{q}s\alpha)$, which are defined as in Eq.~\eqref{eq:elph2} but with respect to the displacement of a single atom $s$ along cartesian component $\alpha$, are written to unformatted binary file ``prefix.epa.k''. Therefore, we introduce a syntax \verb|electron_phonon = 'epc'| in \verb|ph_mat.x| to write the complex number of $\mathcal{M}_{\text{ep}}^{i \to f}(\mathbf{k},\mathbf{q}\nu)$ in Eq.~\eqref{eq:elph2} to formatted file ``prefix.elph''.

\bigskip
\noindent \emph{\bf Acknowledgements}
N.T.H. acknowledges financial support from the Frontier Research Institute for Interdisciplinary Sciences, Tohoku University. R.S. acknowledges JSPS KAKENHI Grants No. JP22H00283. T.Y. and J.Q.H. acknowledge the National Natural Science Foundation of China Grants No. 52031014 and the National Key R\&D Program of China (2022YFA1203901).

\bibliographystyle{elsarticle-num}
%\bibliographystyle{elsarticle-num_SS}
%\bibliography{nguyen.bib}

\end{document}